\documentclass[prb,aps,twocolumn,showpacs,amsmath,amssymb,amsfonts]{revtex4}
\usepackage{graphicx}
\usepackage{color}
\usepackage{bm}

\begin{document}
\title{Electrodynamics of Fulde-Ferrell-Larkin-Ovchinnikov superconducting state} 
\author{ M. Houzet and V. P. Mineev } 
\affiliation{Commissariat \`a l'Energie Atomique,
DSM/DRFMC/SPSMS 38054 Grenoble, France} 
\date{\today}

\begin{abstract}
We develop the Ginzburg-Landau theory of the vortex lattice in clean
isotropic three-dimensional superconductors at large Maki parameter,
when inhomogeneous Fulde-Ferrell-Larkin-Ovchinnikov state is favored.
We show that diamagnetic superfluid currents mainly come from 
paramagnetic interaction of electron spins with local magnetic field,
and not from kinetic energy response to the external field as usual.
We find that the stable vortex lattice keeps its triangular structure as in usual
Abrikosov mixed state, while the internal magnetic field acquires components 
perpendicular to applied magnetic field. Experimental possibilities related 
to this prediction are discussed.
\end{abstract}
\pacs{74.20.Fg, 74.25.Dw, 74.62.Dh}

\maketitle

\section{Introduction}

Orbital and paramagnetic effects are both important in the
suppression of superconducting state. While orbital effect leads to the
formation of Abrikosov vortex lattice below the orbital upper critical
field $H_{c20}\approx \Phi_{0} /2\pi\xi_{0} ^{2} $ in type II superconductors, \cite{Abr} 
paramagnetic effect determines the paramagnetic limit of
superconductivity $H_{p} =\Delta_{0} /\sqrt{2}\mu$,\cite{Clog, Chandr} 
and promotes the tendency to the Cooper pairing with nonzero momentum - so called
Fulde-Ferrell-Larkin-Ovchinnikov state (FFLO).\cite{FF, LO} Here
$\Phi_{0} $ is the flux quantum, $\xi_{0} $ is the superconducting
coherence length, $\Delta_{0} $ is the superconducting gap at zero
temperature and $\mu=g\mu_{B} /2$ is the electron magnetic moment.
 
The interplay of both effects takes place at large enough ratios
$\sqrt{2}H_{c20} /H_{p} =\alpha_{M} $,\cite{GG} called the Maki 
parameter.\cite{Maki}
In this regime, the $(H,T)$-phase diagram of isotropic s-wave superconductor in 
the clean limit was studied by means of a Ginzburg-Landau 
(GL) functional.\cite{Houz}
Critical assumption was made that screening supercurrents are not important,
and the local field in the superconductor was taken equal to the external field.
This is true when the GL parameter $\kappa$ is large enough.
For the clean superconductor, the GL parameter is related to 
the Maki parameter
\begin{equation}
\kappa\approx\frac{\alpha_{M} }{\sqrt{k_{F} r_{e} (m^{*} /m)}},
\label{e1}
\end{equation}
here $k_{F} $ is the Fermi momentum, $r_{e} =e^{2} /mc^{2}$ is classical 
radius of electron and $m^{*} /m$ is the ratio of an effective to the
bare electron mass.  It is clear from this relation that, if the Maki parameter
is large, the GL parameter has even larger value, hence the 
assumption done in Ref.~\onlinecite{Houz} seems reasonable.

If, however, one is interested in the magnetic response of superconductors
in FFLO state, it is necessary to relax this assumption. In the present article, 
we develop the theory of FFLO state in a clean isotropic three-dimensional superconductor,
including the space variations of currents and fields. 

The article is organized as follows. In Section II, we introduce the free energy functional 
including the Zeeman interaction with spatially non uniform magnetic field. Being unimportant 
for the upper critical field determination (Section III) and for the vortex lattice structure 
found in the Section IV, this term is crucial for the current and field distribution 
in FFLO modulated superconducting state (Section V). 
There we find that the internal magnetic field has components perpendicular to applied magnetic field. 
The experimental possibilities related to this theoretical prediction are discussed in Conclusion.

\section{Free energy}

We investigate the phase diagram of a superconductor near the
upper critical field, but far from critical temperature $T_c$. For this purpose,
a GL expansion of the free energy in powers of the order parameter and its 
gradients is possible provided the length scale of variations of the order parameter,
determined by the magnetic length, remains large compared to the
superconducting coherence length. This is indeed the case in the limit of large Maki parameter
when the critical field is mainly determined by paramagnetic depairing effect.

As in Ref. \onlinecite{Houz06}, the free energy density was derived both in frame of Gor'kov\cite{Gor} 
and Eilenberger\cite{Eil} or Larkin-Ovchinnikov\cite{LaOv, Ovc} formalisms:\cite{pauli} 
\begin{eqnarray}\label{eq:functional} 
&&
{\cal F}_s
=
{\cal F}_{n0}+
\frac{\bm{h}^{2}}{8\pi}
+\alpha|\Delta|^{2}+\beta|\Delta|^{4} 
+\gamma|\bm{D}\Delta|^{2}  
\\
&&
+\delta \left [
|\bm{D}^{2} \Delta|^{2} + (2e\bm{h})^{2} 
|\Delta|^{2} -\frac{2e}{3}(\Delta^{*} \bm{D}\Delta + c.c.
)\text{rot}\bm{h}
\right ]
\nonumber \\
&&
+\zeta|\Delta|^{2} |\bm{D}\Delta|^{2} +
\eta[(\Delta^{*} )^{2} (\bm{D}\Delta)^{2}  +
c.c.]
+
\varepsilon (h_z-B)|\Delta|^2.
\nonumber 
\end{eqnarray}
Here, ${\cal F}_{n0}$ is the free energy density
in normal state in absence of magnetic field, $\bm{D}=-i\bm{\nabla}+2e\bm{A}$, 
$\bm{h}=\text{rot}\bm{A}$ is the local internal magnetic field and the coefficients 
in the functional depend both on temperature $T$ and induction $B$ determined  by the spatial average 
$\overline{\bm{h}}\equiv\bm{B}=B\hat{z}$:
\begin{eqnarray}\label{eq:coef}
&&\alpha= N_0 \left(
\ln\frac{T}{T_c}
+2\pi T\Re\sum_{\omega >0}
\left(\frac{1}{\omega}-
\frac{1}{\omega+i\mu B}\right)\right),
\nonumber \\
&&\gamma=\frac{\pi N_0 v_F^2}{12}K_3,\quad
\beta=\frac{\pi N_0}{4}K_3,\quad
\delta=-\frac{\pi N_0 v_F^4}{80}K_5,
\nonumber \\
&&
\zeta=8\eta=-\frac{\pi N_0 v_F^2}{6}K_5,\quad
\varepsilon=-\pi N_0\mu L_2,
\end{eqnarray}
where $N_0=m^*k_F/2\pi^2$ is the normal density of states at Fermi level, $v_F$ is a Fermi velocity, $\omega=\pi T(2\nu+1)$ is a Matsubara frequency, and
\begin{equation}
K_n=2T\Re\sum_{\omega>0}\frac{1}{(\omega+i\mu B)^n},\quad
L_n=2T\Im\sum_{\omega>0}\frac{1}{(\omega+i\mu B)^n}.
\end{equation}

Standard form of GL functional is given by the terms in the first line of 
eq.~(\ref{eq:functional}) only. In the paramagnetic limit, 
transition from normal to uniform superconducting state takes place 
at critical field $B_c(T)$ defined by $\alpha=0$. Along this transition line, 
the coefficients $\beta$ and $\gamma$ become negative 
at $T<T^*\simeq 0.56 T_c$. This signals possible instability toward FFLO
state with spatial modulation of the order parameter $\Delta$, as well as possible change of 
the normal to superconducting phase transition order, at the tricritical point
$(T^*,B^*=B_c(T^*))$. Higher order terms must be retained in 
the functional density (\ref{eq:functional}) to consider these effects, 
as shown in Ref.~\onlinecite{Buz}.

A functional similar to (\ref{eq:functional}), but including arbitrary disorder, 
shape of the Fermi surface and pairing symmetry
of the superconducting state was derived in the limit of 
$\kappa\rightarrow\infty$.\cite{Houz06} At finite value of $\kappa$, 
the coordinate dependent deviation of
magnetic field from external field manifests itself not only in gradient
terms, but also in Zeeman interaction with electron spins. This results in 
the last term in the functional which is absent in Refs.~\onlinecite{Houz, Houz06, Ovc},
and which simply corresponds to local decrease (enhancement) of
the critical temperature $T_c(B)$ when $h_z(\bm{x})>B$ ($<B$). 
Note that the coefficient $\varepsilon$ is proportional to $B$.
Hence, the corresponding term in functional (\ref{eq:functional}) 
is negligibly small in ordinary GL region near $T_c(B\to 0)$

\section{Upper critical field} 

At the second order phase transition between normal and superconducting state,
the magnetic field is uniform, $\bm{h}_{c2}=B\hat{z}$. The linearized gap 
equation obtained from (\ref{eq:functional})
\begin{equation}\label{eq:linear-gap}
\alpha \Delta+\gamma\bm{D}^2\Delta+\delta[(\bm{D}^2)^2+\lambda^{-2}]\Delta=0,
\end{equation}
where $\lambda^{-1} =\sqrt{2eB}$ is inverse magnetic length, is solved with:  
\begin{equation} \label{eq:gapc2}
\Delta=\varphi_{0} (x,y)f(z), \qquad
\bm{D}^{2} \Delta= \left(\frac{1}{\lambda^{2} }+q^{2} \right)\Delta
\end{equation}
where $\varphi_{0} (x,y)$ is the linear combination of Landau wave functions with
level $n=0$ for the particle with charge $2e$ under magnetic field $B$, 
multiplied by exponentially modulated function
along $\hat{z}$-direction, $f(z)=e^{\pm iq z}$. 
(Note that, in principle, higher Landau levels could also be considered. But they 
may be realized only at low temperatures when transition
from normal to superconducting state with lowest Landau level has turned first order
\cite{Houz06}, see Section IV.)

At large Maki parameter, the upper critical field is close to
$B_c(T)$. We note that $\alpha\simeq \varepsilon (B-B_c(T))$
and eq.~(\ref{eq:linear-gap}) then defines a field:
\begin{equation}  \label{eq:bc2q}  
B(q) =B_c (T)
-\frac{\gamma}{\varepsilon}(\lambda^{-2}+q^2)
-\frac{\delta}{\varepsilon}[(\lambda^{-2}+q^2)^2+\lambda^{-4}].
\end{equation}
The upper critical field $B_{c2}$ is found by taking the maximum of $B(q)$
with respect to $q$. Analysis of eq.~(\ref{eq:bc2q}) shows that,
in the presence of orbital effect, the critical temperature $\tilde{T}^*$ below which
FFLO modulation appears, defined by $\gamma+2\delta/\lambda^2=0$, is decreased compared
to its value $T^*$ in absence of orbital effect:\cite{Houz}
\begin{equation}
\tilde{T}^*\simeq T^*-1.2\frac{T_c}{\alpha_M}.
\end{equation}
Namely, at $T>\tilde{T}^*$, the usual superconducting state with $q=0$ appears with critical field
\begin{equation}
B_{c2}(T) =B_c(T)
-\frac{2eB_c(T)\gamma}{\varepsilon}
-\frac{8e^2B_c(T)^2\delta}{\varepsilon}.
\end{equation}
While, at $T<\tilde{T}^*$, the FFLO state appears with finite $q$:
\begin{equation}\label{eq:q0}
\gamma+2\delta(q^2+\lambda^{-2})=0,
\end{equation}
and critical field
\begin{equation}
B_{c2}(T) =B_c(T)
+\frac{\gamma^2}{4\delta\varepsilon}
-\frac{4e^2B_c(T)^2\delta}{\varepsilon}.
\end{equation}

\section{Phase diagram and vortex lattice structure}

In order to determine the structure of the vortex lattice state, 
following variational procedure similar to Refs.~\onlinecite{Abr,deGennes}, consideration of
higher order terms in free energy density (\ref{eq:functional}) is required. 
Just below the upper critical line defined by $B_{c2}(T)$, the magnetic field is partially
screened by supercurrents and we decompose $\bm{h}=\bm{B}+\bm{h}_1$, with 
$B\lesssim B_{c2}$ and $\overline{\bm{h}_1}=0$, and, correspondingly, 
$\bm{A}=\bm{A}_0+\bm{A}_1$. 
At $T>\tilde{T}^*$, conventional Abrikosov state (A state) is realized, and $f(z)=1$.
At $T<\tilde{T}^*$, FFLO state is realized and two possible modulations could appear 
at $B<B_{c2}(T)$: the so-called
FF state with exponential modulation, $f(z)=\exp(iqz)$, and the LO
state with sinusoidal modulation, $f(z)=\sqrt{2}\sin qz$.

By minimizing free energy, we may determine the lattice geometry\cite{Abr} and, 
below $\tilde{T}^*$, the temperature range when FF or LO state
is favored. For this, we write spatial average of the free energy density (
\ref{eq:functional}) in the form:
\begin{equation}\label{eq:freeav}
\overline{{\cal F}_s}
=
\frac{B^2+\overline{\bm{h}_1^2}}{8\pi }
+\overline{{\cal F}_2(\Delta,\bm{A})}
+\overline{{\cal F}_4(\Delta,\bm{A})}
\end{equation}
where ${\cal F}_2$ and ${\cal F}_4$ collect together quadratic and quartic 
terms with respect to $\Delta$, respectively.
By making substitution $\Delta \rightarrow (1+\epsilon)\Delta$ in $\overline{{\cal F}_s}$
and requiring that linear-in-$\epsilon$ terms vanish,\cite{deGennes} we get:
\begin{eqnarray} \label{eq:modulus}
0&=&\overline{{\cal F}_2(\Delta,\bm{A})}
+2\overline{{\cal F}_4(\Delta,\bm{A})}
 \\
&\simeq&
\overline{{\cal F}_2(\Delta,\bm{A}_0)}
+\overline{\bm{A}_1.\frac{\delta{\cal F}_2}{\delta \bm{A}}(\Delta,\bm{A}_0)}
+2\overline{{\cal F}_4(\Delta,\bm{A}_0)}.
\nonumber
\end{eqnarray}
By variation of the free energy with respect to $\bm{A}$, we get the Maxwell equation
relating internal magnetic field to screening currents:
\begin{equation}\label{eq:maxwell}
\frac{1}{4\pi }\text{rot}\bm{h}_1
=
\bm{j}_s
=
-\frac{\delta{\cal F}_2}{\delta \bm{A}}(\Delta,\bm{A}_0),
\end{equation}
up to second order terms $\Delta$. 
We insert Eq.~(\ref{eq:maxwell}) into (\ref{eq:modulus}) and we integrate the second term by part.
Then, eq.~(\ref{eq:modulus}) yields: 
\begin{equation} 
\overline{|\Delta|^2}
=
-\frac{1}{2}
\frac{\overline{{\cal F}_2(\Delta,\bm{A}_0)}/\overline{|\Delta|^2}}
{\left(\overline{{\cal F}_4(\Delta,\bm{A}_0)}-\overline{\bm{h}_1^2}/{8\pi }
\right)/(\overline{|\Delta|^2})^2},
\label{eq:Delta2}
\end{equation}
where r.h.s depends on the structure of order parameter only. 
Near transition, we note that $\overline{{\cal F}_2(\Delta,\bm{A}_0)}
\simeq \varepsilon(B-B_{c2}(T)) \overline{|\Delta|^2}$.
Inserting eq.~(\ref{eq:Delta2}) into (\ref{eq:freeav}), we obtain:
\begin{equation}\label{eq:freered}
\overline{{\cal F}_s}
=
\frac{B^2}{8\pi }
-\frac{\varepsilon^2(B-B_{c2}(T))^2}
{4[\overline{{\cal F}_4(\Delta,\bm{A}_0)}-\overline{\bm{h}_1^2}/{8\pi }]/
(\overline{|\Delta|^2})^2}
.
\end{equation}
The equilibrium vortex lattice structure is thus the one which minimizes the denominator 
of the second term in r.h.s. of eq.~(\ref{eq:freered}).

At large enough GL parameter, the denominator in the second term of eq.~(\ref{eq:freered})
is dominated by its first term.
Noting that the gap function (\ref{eq:gapc2}) obeys the properties:
\begin{equation} 
\label{eq:prop}
(\bm{D}_{\perp} \Delta)^{2} =0,
\qquad
\overline{|\Delta|^{2} |\bm{D}_{\perp} \Delta|^{2}}
=\frac{1}{2\lambda^2}\overline{ |\Delta|^{4}},
\end{equation} 
and using eq.~(\ref{eq:q0}) in FF or LO state, we find:
\begin{equation} \label{eq:F4}
\frac{\overline{{\cal F}_4(\Delta,\bm{A}_0)}}
{(\overline{|\Delta|^2})^2}
=
\pi N_0 \beta_A  \times
\left\{
\begin{array}{ll}
\frac{K_3}{4}-\frac{v_F^2 K_5}{12\lambda^2} & \text{in A state}, \\
-\frac{K_3}{6}+\frac{v_F^2 K_5}{24\lambda^2} & \text{in FF state}, \\
\frac{K_3}{36}-\frac{v_F^2 K_5}{48\lambda^2} &\text{in LO state}. 
\end{array}
\right.
\end{equation}
where
\begin{equation}
\beta_{A} =\frac{\overline{|\varphi_0|^{4} }}{(\overline{|\varphi_0|^{2} })^{2} }
\end{equation}
is the Abrikosov parameter. \cite{Abr}
Comparison of the expressions in r.h.s. of (\ref{eq:F4}) yields \cite{Houz, Houz06} that
FF state is realized in the temperature range defined by
$9/28>\lambda^2K_3/v_F^2K_5>3/10$, that is, 
$T_0\equiv \tilde{T}^*-0.08T_c/\alpha_M<T<\tilde{T}^*$. LO state is realized
below these temperatures, and the transition from normal to LO
state changes its order at $\lambda^2K_3/v_F^2K_5>3/4$, that is
$T<T_1\equiv\tilde{T}^*-2T_c/\alpha_M$.

At all temperatures, free energy (\ref{eq:freered}) is minimized when $\beta_A$ is 
minimal, that is for triangular vortex lattice (with $\beta_A=1.1596$). 
Corresponding phase diagram is shown qualitatively in the Figure.

\begin{figure}
\includegraphics[width=75mm,angle=-90]{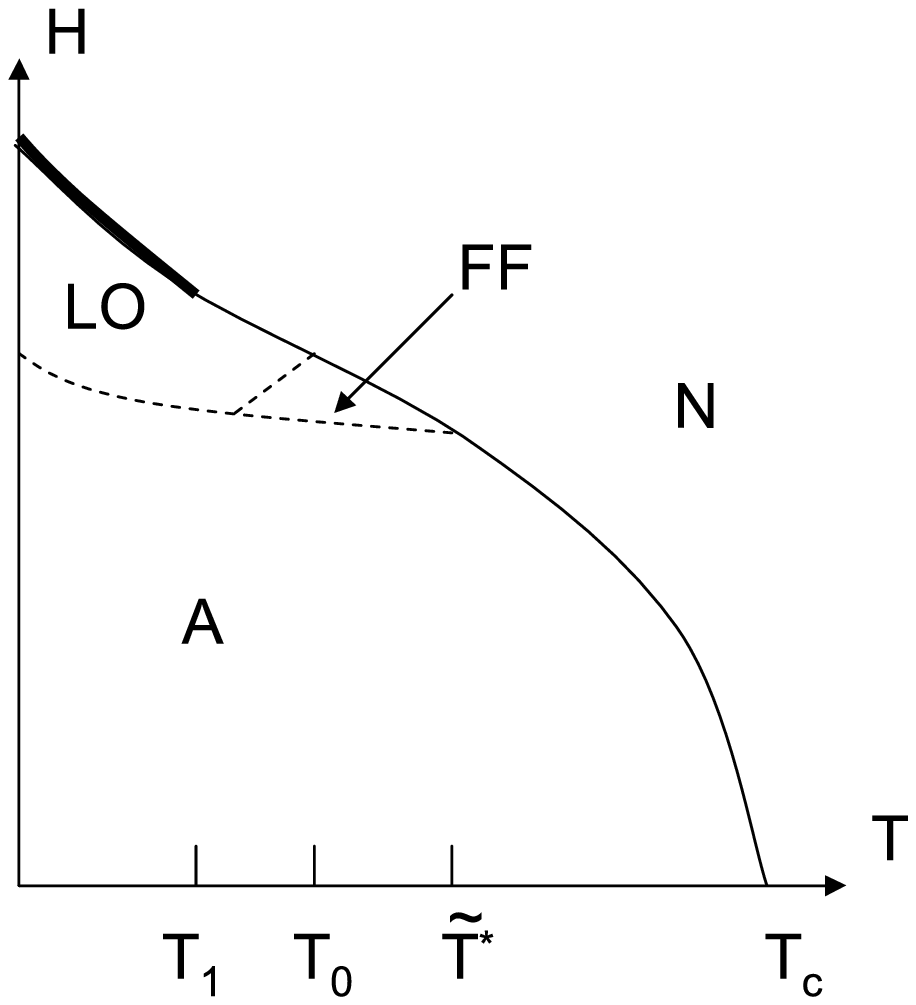}
\caption{
Qualitative phase diagram of
a clean three-dimensional superconductor with
large Maki parameter. Thin (thick) line is for second (first) order
transition. Transitions shown with dashed lines are not discussed 
in the present work. 
}
\label{fig}
\end{figure}

\section{Electrodynamics of vortex states}

To consider the situation at finite GL parameter
we need to evaluate the term $\overline{\bm{h}_1^2}$ in eq.~(\ref{eq:freered})
where $\bm{h}_1$ solves the Maxwell equation (\ref{eq:maxwell}) with:
\begin{subequations}
\begin{equation} \label{eq:j}
\bm{j}_s=\bm{j}_\text{kin}+ \bm{j}_\text{Z},
\end{equation}
\begin{eqnarray}
\label{eq:j_k}
\bm{j}_\text{kin}&=&-2e\left\{
(\gamma\Delta+2\delta\bm{D}^{2} \Delta)(\bm{D}\Delta)^{*} \right. \\
&&\left.
-\frac{\delta}{3} \text{rot}\,\text{rot}[\Delta({\bf D}\Delta)^{*}]+c.c.
\right\}+
8e^2\delta\bm{B} \times\nabla |\Delta|^{2},\nonumber
\\
\label{eq:j_p}
\bm{j}_\text{Z}&=&-\varepsilon~ \text{rot}(|\Delta|^{2} \hat z).
\end{eqnarray}
\end{subequations}
Here, $\bm{j}_\text{kin}$ originates from usual superconducting kinetic energy 
response to the spatially varying magnetic field. 
The Zeeman current $\bm{j}_\text{Z}$ arises from the interaction energy of 
the superconducting diamagnetic correction to the normal-metal
paramagnetic moment, with the spatially varying magnetic field ${\bm{h}_1}$.

Making use of the form (\ref{eq:gapc2}) for the gap and its property
(at Landau level $n=0$)
\begin{equation}
\Delta^* (\bm{D}_\perp\Delta) +c.c.=\text{rot}(|\Delta|^{2} \hat{z}),
\end{equation}
we find:
\begin{subequations}
\begin{equation} 
\bm{j}_s=\bm{j}_{s,\perp}+\bm{j}_{s,\parallel},
\end{equation}
\begin{equation}
\bm{j}_{s,\perp} 
=
-\text{rot}
\left[
\left(\tilde{\epsilon}+\frac{2e\delta}{3}\nabla^2\right)
|\Delta|^2\hat{z}
\right],
\end{equation}
where $\tilde{\epsilon}=\epsilon+2e(\gamma+4\delta/\lambda^2)$
in A state, and, with use of eq.~(\ref{eq:q0}), 
$\tilde{\epsilon}=\epsilon+4e\delta/\lambda^2$
in FF and LO states. And for longitudinal component, again with use of eq.~(\ref{eq:q0}),
we have in the lowest order in $|\Delta|^2$: 
\begin{eqnarray}
\label{eq:26}
\bm{j}_{s,\parallel}
&=&
\left\{
\begin{array}{ll}
0&\text{in A and LO states}, \\
-4e\frac{q\delta}{3}\nabla^2|\Delta|^2\hat{z}
&\text{in FF state.}
\end{array}
\right.
\end{eqnarray}
\end{subequations}

Let us note that in A state or LO state,
the supercurrents flow in planes perpendicular to the external field only, while
in FF state, they also flow in parallel direction. 
Close to the critical temperature $\tilde{T}^*$, the simple estimation shows that
\begin{equation}    
\left |\frac{\bm{j}_\text{kin}}{\bm{j}_{Z}}\right |~\approx\frac{1}{\alpha_M^{2}.}
\end{equation}
Hence, we may neglect $\bm{j}_\text{kin}$ in perpendicular component of current.
In A and LO states, the Maxwell equation (\ref{eq:maxwell}) acquires the following form
\begin{equation}\label{eq:M}
-\frac{\text{rot}\bm{h}_1}{4\pi}
\simeq 
\varepsilon~
\text{rot}|\Delta|^2\hat{z}.
\end{equation}
In FF state, we may keep the small term corresponding to the 
parallel-current component and we get:
\begin{equation}\label{eq:M_F}
-\frac{\text{rot}\bm{h}_1}{4\pi}
\simeq 
\varepsilon~
\text{rot}|\Delta|^2\hat{z}+\frac{4e\delta q}{3}\nabla^2|\Delta|^2\hat z.
\end{equation}

Let us remind that in conventional Abrikosov state in vicinity of critical
temperature, where coefficient $\gamma$ is positive, the supercurrents
are diamagnetic: that is, they create a magnetic moment directed
opposite to the external field. 
The value of $\varepsilon$ is always positive, hence as in the Abrikosov case 
the orbital currents in FF and LO modulated states are also diamagnetic despite their Zeeman origin.

Let us now determine the distribution of fields in vortex state.
The component $\bm{h}_1$ of the magnetic field is periodic and is found
from Maxwell equations (\ref{eq:M}) or (\ref{eq:M_F}), and
\begin{equation}
\text{div}\bm{h}_1=0
\end{equation} 
together with the condition $\overline{\bm{h}_1}=0$. 
In A state, these equations are solved with:
\begin{equation} \label{eq:A}
-\frac{\bm{h}_1}{4\pi}
=
\varepsilon
(|\Delta|^2-\overline{|\Delta|^2})
\hat{z}.
\end{equation}
In FF state, they are solved with
\begin{equation}
\label{eq:I}
-\frac{\bm{h}_1}{4\pi}
= 
\varepsilon
(|\Delta|^2-\overline{|\Delta|^2})
\hat{z}
+\frac{4e\delta q}{3}(\hat z\times \bm{\nabla}|\Delta|^2).
\end{equation}
In LO state, we search for the solution in following form:
\begin{equation}
\label{eq:II}
-\frac{\bm{h}_1}{4\pi}
=\varepsilon
(|\Delta|^2-\overline{|\Delta|^2})
\hat{z}+2q\bm{\nabla}[\chi(x,y)\sin 2qz],
\end{equation} 
and we find that $\chi(x,y)$ is an auxiliary function which solves:
\begin{equation}
\label{eq:chi}
-\bm{\nabla}^2_\perp \chi
+4q^2\chi
=\varepsilon|\varphi_0|^2
.
\end{equation}
In FF and LO  states, in contrast with the situation in Abrikosov vortex
state, the internal field has components perpendicular
to applied magnetic field. 

Evaluating $\overline{\bm{h}_1^2}$ in A and FF state, we get 
(neglecting the transverse component of field in FF state):
\begin{equation}\label{eq:varh1}
\frac{\overline{\bm{h}_1^2}}{8\pi(\overline{|\Delta|^2})^2}=
2\pi\varepsilon^2(\beta_A-1).
\end{equation}
In LO state, we get:
\begin{equation} \label{eq:h1LO}
\frac{\overline{\bm{h}_1^2}}{8\pi(\overline{|\Delta|^2})^2}=
2\pi\varepsilon^2\left(\frac{3}{2}\beta_A-1+
\frac{2q^2\overline{\chi|\varphi_0|^2}}{\varepsilon}\right).
\end{equation}
A rough evaluation of the last term in (\ref{eq:h1LO}) can be done by
estimation of $-\bm{\nabla}^2_\perp \chi \approx \lambda^{-2} \chi $ in (\ref {eq:chi}).
Consequently similar to (\ref{eq:varh1})
we obtain\cite{Yang}:
\begin{equation}\label{eq:avh12}
\frac{\overline{\bm{h}_1^2}}{8\pi(\overline{|\Delta|^2})^2}=
2\pi\varepsilon^2(C\beta_A-1),
\end{equation}
where in A and FF states $C=1$ and for LO state
\begin{equation}
C_{LO}=\frac{3}{2}+\frac{2q^2\lambda^2}{1+4q^2\lambda^2}.
\end{equation}

Inserting the result for $\overline{\bm{h}_1^2}$ in (\ref{eq:freered})
and using eq.~(\ref{eq:F4}),
we can present the free energy in usual form:\cite{Abr,deGennes}
\begin{equation}\label{eq:freered2}
\overline{{\cal F}_s}
=
\frac{B^2}{8\pi }
-\frac{(B-B_{c2}(T))^2}
{8\pi[1+\beta_A(2\kappa_{\text{eff}}^2-C)]}
,
\end{equation}
with effective, temperature-dependent, GL parameter 
$\kappa_{\text{eff}}$ defined by
\begin{equation}\label{eq:keff}
\kappa_{\text{eff}}^2
=
\frac{
	N_0}
{4 \varepsilon^2}
\times
\left\{
\begin{array}{ll}
\frac{K_3}{4}-\frac{v_F^2 K_5}{12\lambda^2} & \text{in A state}, \\
-\frac{K_3}{6}+\frac{v_F^2 K_5}{24\lambda^2} & \text{in FF state}, \\
\frac{K_3}{36}-\frac{v_F^2 K_5}{48\lambda^2} &\text{in LO state}. 
\end{array}
\right.
\end{equation}

We note that the form (\ref{eq:freered2}) of the free energy requires
$\kappa_{\text{eff}}^2>0$, that is $T>T_1$. Below $T_1$, normal to FFLO
state transition becomes of the first order and requires higher order terms
in the gap to be retained in eq.~(\ref{eq:functional}). 
Moreover, at $T>T_1$, in analogy with type I/type II superconductors, the 
free energy (\ref{eq:freered2}) is indeed minimized with vortex lattice state 
only if $\kappa_{\text{eff}}>\sqrt{C/2}$. The situation in vicinity of the point 
$\kappa_{\text{eff}}=\sqrt{C/2}$ requires special investigation similar to 
Ref. \onlinecite{Luk} at $T \to T_c$.
Evaluating $\kappa_{\text{eff}}$ at 
temperature $\tilde{T}^*$ corresponding to the
tricritical point for N/A/FF states, we find:
\begin{equation}
\kappa_\text{eff}^2
\simeq
\frac{\kappa^2}{\alpha_M^3}.
\end{equation}
The theory of vortex lattice in FFLO state thus applies at:
\begin{equation}
\frac{\kappa^2}{\alpha_M^3}\sim 
\frac{1}{k_F  r_e}
\frac{m}{m^*}
\frac{1}{\alpha_M}
\gtrsim
1.
\end{equation}

We now determine the diamagnetic response. Applied magnetic field is found 
from thermodynamic relation:
\begin{equation}
H=4\pi\frac{\partial {\cal F}_s}{\partial B}.
\end{equation}
From relation $B=H+4\pi M$, we obtain the magnetization induced in superconducting state
at a given applied magnetic field
\begin{equation}
M(H)=-
\frac{1}{4\pi}
\frac{B_{c2}-H}
{(2\kappa_\text{eff}^2-C)\beta_A}.
\end{equation}
We note that the derivative of induced magnetization in superconducting state 
with respect to applied magnetic field varies with the temperature along the upper
critical line $(T,H_{c2}(T))$. Its most peculiar features are: 1) it increases
abruptly close to temperature $T_0$, as $C=1$ at $T>T_0$ when the transition 
is from N to FF state, and $C> 3/2$ at $T<T_0$ when the transition 
is from N to LO state; 2) it diverges at $T<T_1$ when transition into LO state 
becomes of the first order. Feature 1) takes place close to the triple point for
coexistence of N, FF and LO state, and it is related to the nature of FF/LO transition
which is of the first order\cite{Houz}.

The determination of magnetic field distribution is also important for 
nuclear magnetic resonance (NMR). Due to different field distributions (\ref{eq:A}),
(\ref{eq:I}), and (\ref{eq:II}) in A, FF, and LO states, respectively, we may expect distinct 
NMR line shapes in each of theses states.

\section{Conclusion}

The experimental search for FFLO state is not so easy due to the absence 
of a particular feature distinguishing it from the ordinary Abrikosov mixed state.
We already pointed out peculiarities of the magnetization and NMR in FFLO state. 
More importantly, the solutions (\ref{eq:I}) and (\ref{eq:II}) 
for the field distribution
demonstrate that the internal field has components perpendicular
to applied magnetic field both in FF and LO states, in contrast with the situation in Abrikosov vortex
state. The transverse component of the oscillating field in FF state is negligibly small in comparison with longitudinal oscillating component, while both components of oscillating field have comparable value in LO state. 

The effect  could be revealed experimentally by 
means of small-angle scattering of neutrons {\it polarized} parallel to the external field. Namely, the transition to the
LO (and possibly to FF) state should manifest itself by the strong increase of scattering with neutron spin flip. Another possibility to reveal LO and FF state is related with application of $\mu$SR technique
by making measurements of relaxation rate of precessional motion of muon spins polarized along the external field direction.

Finally, it should be noted that the effect of appearance of space oscillating transverse field component in LO and FF states  found here for the isotropic s-wave superconductor has model independent character and will also be present in anisotropic materials with a different type of superconducting pairing.

\acknowledgments

We would like to thank A. Buzdin for careful reading of the manuscript and J. Flouquet for strong interest in the present work.


\begin{thebibliography}{99}    

\bibitem{Abr}A.A.Abrikosov, Zh.Eksp.Teor.Fiz.{\bf 32}, 1442 (1957)
[Sov.Phys.JETP {\bf 5}, 1174 (1957)].



\bibitem{Clog} A.M.Clogston, Phys.Rev.Lett. {\bf 9}, 266
(1962).

\bibitem{Chandr} B.S.Chandrasekhar, Appl.Phys.Lett.  {\bf 1}, 7 
(1962).
 
\bibitem{FF} P.Fulde and R.A.Ferrell, Phys.Rev.  {\bf 135}, A550 
(1964).

\bibitem{LO} A.I.Larkin and Yu.N.Ovchinnikov,  Zh.  Eksp.Teor.Fiz.  
{\bf 47}, 1136 (1964) [Sov. Phys. JETP {\bf 20}, 762 (1965)].


\bibitem{GG} L.W.Gruenberg and L.Gunther,  Phys.Rev.Lett.  {\bf 16}, 
996 (1966)

\bibitem{Maki} K.Maki, Physics {\bf 1}, 127 (1964). 


\bibitem{Houz}M.Houzet, A.I.Buzdin,  Phys.Rev.B {\bf 63}, 184521 
(2001).

\bibitem{Houz06}M.Houzet, V.P.Mineev, Phys.Rev.B {\bf 74},
144522 (2006).

\bibitem{Gor}L.P.Gor'kov, JETP {\bf 36}, 1918 (1959) [Soviet Phys. JETP {\bf 9}, 1364 (1959)].

\bibitem{Eil} G. Eilenberger, Z.Phys.{\bf 214}, 195 (1968).

\bibitem{LaOv} A. I. Larkin and Yu. N. Ovchinnikov, Zh.  
Eksp.Teor.Fiz.  
{\bf 55}, 2262 (1968) [Sov. Phys. JETP {\bf 28}, 1200 (1969)].

\bibitem{Ovc} Yu. N. Ovchinnikov,  Zh.  Eksp.Teor.Fiz.  
{\bf 115}, 726 (1999) [JETP {\bf 88}, 398 (1999)].

\bibitem{pauli} Here, we assume that Pauli spin susceptibility 
is negligible: 
$\chi_P=2\mu^2N_0\sim (m^*/m)k_Fr_e\ll 1$, and magnetic permeability is 
taken equal to 1.

\bibitem{Buz} A.I.Buzdin and H.Kachkachi, Phys.Lett.  A{\bf 225}, 341
(1997).


\bibitem{deGennes}P.G.de Gennes, {\it {Superconductivity of Metals and Alloys}}
(W.A.Benjamin, INC.,New-York - Amsterdam, 1966).

\bibitem{Yang} A more precise evaluation of 
$\overline{\chi|\varphi_0|^2}$ can be done with the method presented in 
[K. Yang and A. H. MacDonald, Phys. Rev. B \textbf{70}, 094512 (2004)].

\bibitem{Luk} I. A. Luk'yanchuk, Phys.Rev.B {\bf 63}, 174504 (2001).

\end{thebibliography}
\end{document}